\documentclass[twocolumn,showpacs,preprintnumbers,amsmath,amssymb]{revtex4}

\usepackage{graphicx}
\usepackage{dcolumn}
\usepackage{bm}

\begin{document}

\preprint{APS/123-QED}

\title{Calibration of second-order correlation functions for non-stationary sources \\ with a multi-start multi-stop time-to-digital converter}

\author{Wonshik Choi, Moonjoo Lee, Ye-ryoung Lee, Changsoon Park, Jai-Hyung Lee, and Kyungwon An}
\email{kwan@phya.snu.ac.kr}
\address{School of Physics, Seoul National University, Seoul, 151-742, Korea}
\author{C.\ Fang-Yen, R.\ R.\ Dasari, and M.\ S.\ Feld}
\address{G.R.Harrison Spectroscopy Laboratory, Massachusetts Institute of Technology, Cambridge, MA 02139 }

\date{\today}

\begin{abstract}
A novel high-throughput second-order-correlation measurement system
is developed which records and makes use of all the arrival times of
photons detected at both start and stop detectors. This system is
suitable particularly for a light source having a high photon flux
and a long coherence time since it is more efficient than
conventional methods by an amount equal to the product of the count
rate and the correlation time of the light source. We have used this
system in carefully investigating the dead time effects of detectors
and photon counters on the second-order correlation function in the
two-detector configuration. For a non-stationary light source,
distortion of original signal was observed at high photon flux. A
systematic way of calibrating the second-order correlation function
has been devised by introducing a concept of an effective dead time
of the entire measurement system.
\end{abstract}

\maketitle

\section{Introduction}

A second-order correlation function is an intensity-intensity
correlation function, having information on both photon statistics
and dynamics of the light generation process of a light source. It
was first introduced by Hanbury Brown and Twiss in order to measure
the angular separation of binary stars~\cite{HBT}, and later it was
applied to property measurement of various light sources such as
measuring the coherence time of thermal light~\cite{Morgan1966},
getting information on the nature of scatterers~\cite{Cummins1970}
and surveying the correlation properties of laser light near laser
threshold~\cite{Arecchi1966}. More recently, it was used in probing
the nonclassical nature of light such as
antibunching~\cite{Kimble-PRL77} and sub-Poissonian photon
statistics~\cite{Short-PRL83}.

Much effort has been made in devising a precise and efficient
apparatus to measure the second-order correlation function. The
first successful time-resolved measurement was done by using a
single detector, a single variable delay generator and a coincident
circuit to measure the coherence time of low-pressure gas discharge
in a single $^{198}Hg$ isotope ~\cite{Morgan1966}. This technique
has a limitation in getting the correct correlation function near
zero time delay due to the imperfectness of the detector such as
spurious multiple pulse emission and incapability of detection for a
finite amount of time just after detecting real photons. These
effects are referred to as after-pulsing and dead time effects,
respectively.

To overcome this limitation, a two-detector configuration has been
adopted in which a light beam is divided into two parts and two
photodetectors are used to record photons arrived at each detector
~\cite{Scarl1966,Phillips1967,Davidson1968,Davidson1969}. Unlike the
single-detector configuration, the spuriously emitted photons and
dead photons at one detector are completely uncorrelated to the
photons detected on the other detector so that the contribution of
after-pulsing and dead time effects are equally spread over the
entire measurement time. This allows measurements to be extended
down to the zero time delay.

More sophisticated correlators, which made use of multiple time
delays, were developed which digitized the time interval between a
{\em start} pulse from one photodetector and multiple {\em stop}
pulses from the other photodetector at a time~\cite{Cummins1977,
Swinney1983, Pike1986}. The number of pulse pairs corresponding to a
given delay is registered on a corresponding counter and the result
is proportional to the second-order correlation function. Such a device is
called a multi-stop time-to-digital converter (MSTDC). This method
is more efficient than previous method owing to the multiple delay
generators.

Even though a two-detector configuration can effectively remove the
artifacts caused by the detector imperfectness, this benefit is true
only for a stationary source, the intensity of which is independent
of time. Since the probability of spurious emissions or losing
photons due to the detector dead time depends on the intensity, the
measured intensity profile can be greatly distorted for a
non-stationary source. In other words, the spurious emissions or
lost photons are not equally spread over the entire measurement time
but dependent on time. This can cause distortion in the second-order
correlation measurement.

In the present study, we have carefully investigated the limitation
of the two-detector configuration for non-stationary sources and
have devised a systematic way of calibrating the second-order
correlation function for the first time to our knowledge. For this
study, we have developed a novel second-order correlation
measurement system which records all the arrival times of photons
detected at start and stop detectors and make use of all the photons
detected at the start detector as triggers. There are no waste of
photons at the start detector and thus our system can be much more
efficient than MSTDC for a light source having a high photon flux
and a long coherence time.

\section{Experimental Apparatus}

\begin{figure}[b]\centering
\includegraphics[width=3.3in]{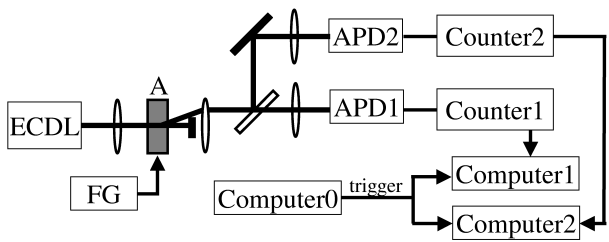}
\caption{Schematic of experimental setup. ECDL:Extended cavity diode
laser(2010M:Newport), A:Acousto-optic modulator(Isomat 1206C),
FG:Function generator(DS345, Stanford Research System), APD1,2:
Avalanche photodiode(SPCM-AQR-13, PerkinElmer), Counter1,2: NI 6602
counter/timing board. Counter/timing boards are installed in
computer 1 and computer 2 separately, and are simultaneously armed
by a trigger signal generated from computer 0.} \label{setup}
\end{figure}

A schematic of our second-order correlation measurement system is
shown in Fig.~\ref{setup}. Photons are detected by two avalanche
photodiodes (APDs, model SPCM-AQR-13 by PerkinElmer). One detector
(APD1) serves as a start detector while the other (APD2) as a stop
detector. Each APD has a dark count rate less than 150 cps and a
dead time of about 50 ns and generates electrical pulses whenever
the photons are detected with detection efficiency of about 50$\%$.
APDs are electrically connected to the counter/timing boards
(Counter 1, 2) installed in two computers (Computer 1, 2).

Relatively low-cost and commercially available counter/timing boards
(PCI-6602 by National Instruments) are used to record the arrival
times of the electrical pulses and to store them to the computers.
Each board has its own internal clock of 80 MHz so that the time
resolution is 12.5ns. Two independent boards installed in separate
computers are used to prevent the crosstalks between boards or
computers.

The internal clock of each counter/timing board has a limited
accuracy and also has a drift of 50 ppm as the surrounding
temperature changes. The accurate frequency of the internal clock in
each board has been calibrated by counting the arrival times of
reference pulses from function generator(DS345, Stanford Research
System). The frequency difference between boards is typically
several tens of ppm and is accounted for in getting absolute arrival
times from the measured ones.

To make the two counters start to count at the same time, an
additional control computer (computer 0) is used to generate a
trigger signal to simultaneously arm the counters. It has a board
with analog outputs and digital inputs/outputs (NI6703, National
Instruments), which can send a TTL signal as a trigger.

All the arrival times of photons detected on both detectors are the
relative times with respect to the same origin defined by the
trigger. All the detected photons on one detector therefore can be
used as start pulses with respect to those of the other detector.
For this reason, we call our apparatus a {\em multi-start multi-stop
time-to-digital converter} (MMTDC) compared to the conventional {\em
multi-stop time-to-digital converter} (MSTDC).

MMTDC makes use of all the photons detected at a start detector
whereas the MSTDC makes use of a single photon detected at a start
detector as a trigger and measures the relative arrival times for a
time interval which is several times (let us say $n$ times) longer
than a correlation time $T_c$ of a light source. It starts over and
repeats the next measurement cycle using another single photon
detected at the start detector after $n T_c$. If the incoming photon
flux to the start detector is $\gamma$, only one photon out of $n
\gamma T_c$ photons is used in the measurement. Therefore, our MMTDC
has an efficiency $n \gamma T_c$ times higher than that of MSTDC.

MMTDC is specially advantageous for a light source which has a high
photon flux and long correlation time but has a limited operation
time with its second-order correlation signal embedded in a large
background. The microlaser ~\cite{An-PRL94, Wonshik04} was a good
example to fit this category. It had an output photon flux of about
3 Mcps and a correlation time of about 10 $\mu s$ such that our new
method was about 30 times more efficient than that of MSTDC. Because
of limited oven life time, full time measurement could give a
signal-to-noise ratio of about 3 even when MMTDC was used. We could
have obtained a signal-to-noise ratio of only 0.55 if we had used
the conventional MSTDC method, only to fail to resolve the signal. 

Due to limitation in computer memory, the number of arrival times
recordable at a time is limited by about one million counts in our
MMTDC setup. To get an enough number of data, measurements should be
done in a sequential way. All computers are connected by ethernet
connections in such a way that they can send and receive messages
among themselves. The counting computers (computer 1 and 2) send a
message to the control computer to notify the end of counting
whenever they complete a specified number of counting and recording.
After checking that the control computer has received the message,
both counting computers prepare a next measurement. When the control
computer receives the message, it sends triggers to the counting
computers to initiate counting again. The number of sequences is
determined so as to get an enough signal-to-noise ratio.

To get the second-order correlation function, a histogram is
constructed for the time differences between all possible pairs made
of one of the photon arrival times at start APD and another one at
stop APD. To save calculation time, only the pairs of photons the
time difference of which are within a certain time window, typically
chosen 10 times larger than the correlation time of the source, are
included in the calculation. The second-order correlation function
can then be obtained from the normalization of this histogram by the
averaged histogram value for much longer delay times than the
correlation time.

\section{Experiments with stationary sources}
We have tested our MMTDC system using an extended cavity diode laser
(2010M, Newport) operating 
far above 
threshold as a test
source. The photon statistics of its output is supposedly Poissonian
and the second-order correlation function is thus unity for
all delay times. Photon flux was 3 Mcps for each detector and the
number of sequences was 1,000 with 500 kilo counts per sequence.

\begin{figure}[t]\centering
\includegraphics[width=3.3in]{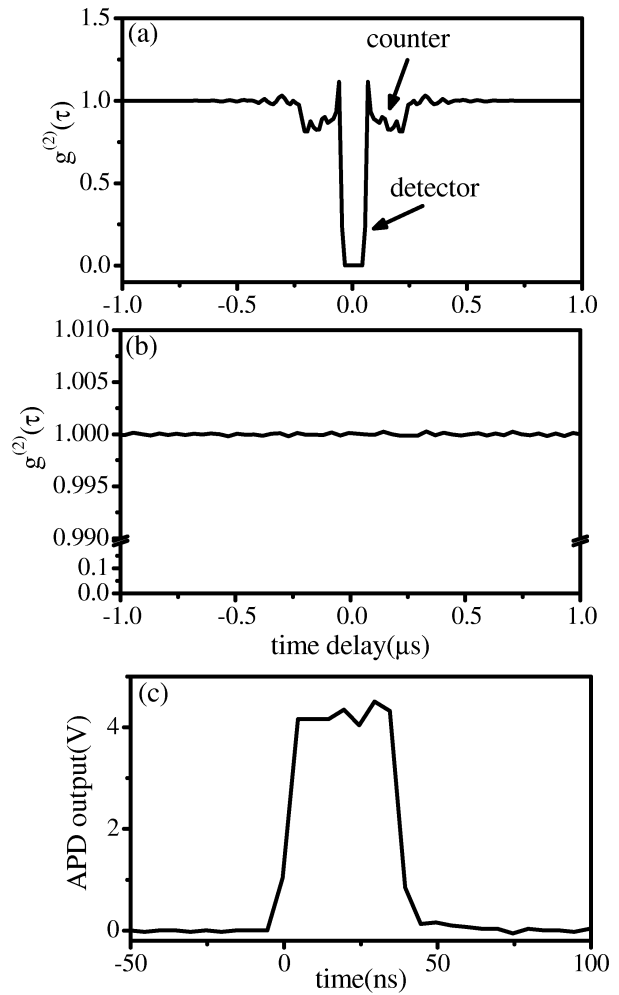}
\caption{(a) Second-order correlation function for a Poissonian
light source measured in the single-detector configuration. (b) The
same measured in the two-detector configuration. (c) Shape of a
single-photon pulse from APD.} \label{DeadTimeDistribution}
\end{figure}

A correlation function obtained from the photons measured at a {\em
single} detector exhibits the effect of detector dead time and
after-pulsing. A figure~\ref{DeadTimeDistribution}(a) shows a
measured result. Two types of dip below 1 appear and the one near
the zero time delay corresponds to the detector dead time. This can
be confirmed from the output pulse shape of APD shown in
Fig.~\ref{DeadTimeDistribution}(c). The full width of the pulse was
measured to be about 50 ns. The stop photons following a given start
photon within this time window are completely ignored and thus a dip
with a depth of unity appears.

Another dip is extended to 250 ns and its depth is about 0.25, which
means that the following stop photons are partially ignored with a
probability of 25\%. This dip originates from the counter/timing
board. Usually counter electronics also have dead times since it
takes a finite amount of time to record measured arrival times. In
our case, since there is no onboard memory in the counter/timing
board, it has to transmit the arrival times to the computer memory
through DMA (Direct Memory Access). Since the data transfer rate
through DMA is limited by 100 MBps, some counts can be missed if the
time interval between photons is too short to be transferred. The
probability to miss a count depends on the time interval between
successive photons. We call this loss of counts the {\em dead time
effect of the counter} as an analogy to the detector dead time.

The correlation function obtained from the photons detected at {\em
two} detectors is shown in Fig.~\ref{DeadTimeDistribution}(b). The
effect of the detector and counter dead time is completely removed
and the value of the correlation was very close to unity for all
delay times, as expected. Since the number of photon pairs per bin
amounted to $10^8$, the standard deviation from unity was only
$10^{-4}$. Utilizing all the photons detected at start detector
helped to reduce the background noise greatly.

\section{Experiments with Non-Stationary Sources}

Using our MMTDC second-order correlation measurement system, we have
measured the second-order correlation function for a non-stationary
light source. The output beam from an extended cavity diode laser
was modulated by an acousto-optic modulator. The amplitude of a
driving RF (radio frequency) field to the acousto-optic modulator
was sinusoidally modulated using a function generator so that the
intensity of the first-order diffracted beam was sinusoidally
modulated. Its functional form can be written as $I(t)=a \sin(\omega
t+\phi)+b$, where $a$ and $b$ have the units of count per second for
the photon counting measurement. Since the response time of AOM is
measured to be 130 ns, the modulation frequency can be safely set up
to 1 MHz. For the present experiments, the modulation frequency was
set as 100 kHz.

\begin{figure}[b]\centering
\includegraphics[width=3.3in]{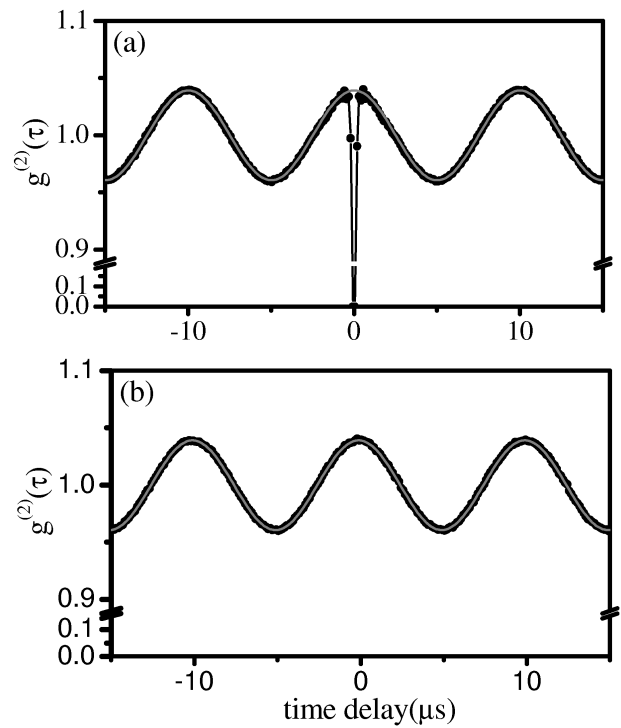}
\caption{Typical result of the second-order correlation function
for a sinusoidally modulated light source measured in (a)
single-detector configuration and (b) two-detector
configuration.}
\label{EliminationDip}
\end{figure}

The normalized second-order correlation function for a classical
source, applied to the intensity of electromagnetic field $I(t)$, is
given by
\begin{equation}
g^{(2)}(\tau)=\frac{\langle I(t)I(t+\tau)\rangle}{\langle I(t)
\rangle ^2},
\end{equation}
where $\langle \ldots \rangle$ denotes a time average. The
second-order correlation function for the sinusoidally modulated
source is thus calculated to be
\begin{equation}\label{second_order}
g^{(2)}(\tau)=1+\frac{a^2}{2 b^2}\cos\omega \tau.
\end{equation}
Figure \ref{EliminationDip}(a) and (b) show measured second-order
correlation functions. Mean count rates for individual detectors
were about 0.6 Mcps and 1000 sequences of measurements were done
with each sequence counting 300 kilo counts for each detector. It
took about 1000 seconds including sequencing procedure. The adjacent
points were added up and thus the time resolution was 125 ns.

Figure \ref{EliminationDip}(a) is the second-order correlation function
obtained from photons detected at a single detector and thus shows a
sharp dip near time delay zero. The dip results from the dead times
of the detector and the counter. On the other hand, Fig.\
\ref{EliminationDip}(b) is obtained from the photons detected on two
detectors, APD1 and APD2. The central dip is completely eliminated,
as in the case of stationary light sources. The normalized shot
noise is only about 0.06\% due to the large mean counts per bin of
about 2.8 million. The contrast ratios $a^2/2b^2$ of single- and
two-detector configurations are almost the same.

\subsection{Dead Time Effect on Non-Stationary Intensity}

\begin{figure}[b]\centering
\includegraphics[width=3.3in]{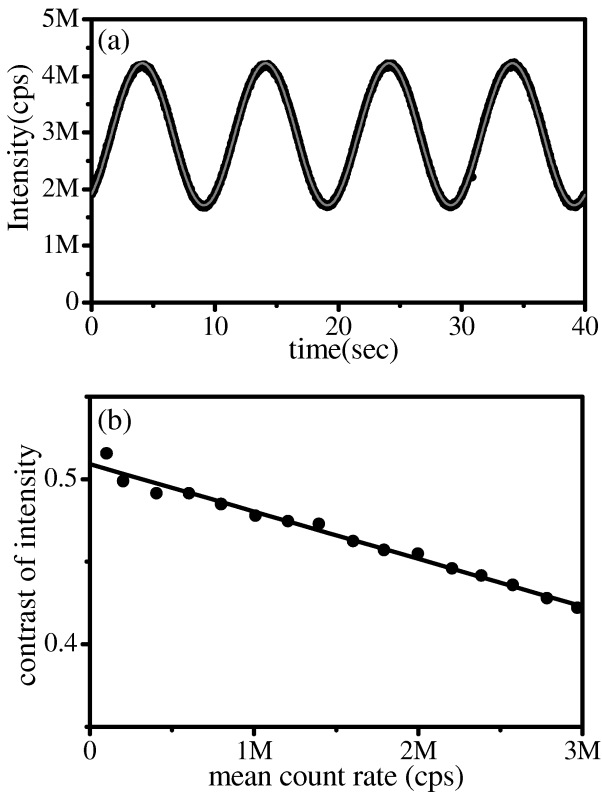}
\caption{(a) Intensity profile of a sinusoidally modulated light
source measured by the photon counting method. (b) Contrasts of the
intensity. Circular dots: measured contrast ratio from fitting the
intensity, line: theoretical calculation including the effect of
detector dead time of 56 ns.} \label{ContrastIntensity}
\end{figure}

Even though the two-detector configuration can eliminate the
detector dead time effect near the zero time delay, the detector
dead time still can affect the correlation measurements for
non-stationary sources when the detector counting rate becomes
comparable to the inverse of the detection dead time. Since the
probability to miss photons is dependent on the intensity, the
measured time-varying intensity profile can be distorted.

We used photon counting method to measure the time-dependent
intensity. Since our MMTDC was based on the photon counting method,
we used photon counting technique rather than photo current 
measurement for a consistent quantitative relation between the intensity
profile and the second-order correlation function. In order to obtain enough
number of counts per counting bin for intensity measurement, which
has to be done in a fashion of single shot measurement, we used a
slow intensity modulation frequency of 0.1 Hz for a given incoming
photon flux of around 1 MHz.

Photons were counted for every counting bin of 0.01 sec and the
counted numbers were transferred at the end of each bin. Since the
dead time of the counter occurs when the time interval between
successive data transfers is shorter than 250ns, this measurement
should be free from the counter dead time effect. At 1 MHz counting
rate, the mean counts $N$ per bin was $10^4$ counts and thus the
normalized noise $\sqrt{N}/N$ was 0.01, which means we could resolve
an intensity modulation whose contrast was as small as 0.01.

We repeated the measurement for various mean count rates $b$ while
keeping the contrast of the intensity $a/b$ constant. To do so, we
fixed the radio frequency field driving the acousto-optic modulator
and varied the mean intensity using neutral density filters. The
intensities measured in this way were well fitted by sinusoidal
functions as shown in Fig.\ \ref{ContrastIntensity}(a). The contrast
ratio in Fig.\ \ref{ContrastIntensity}(a) is about $0.422$, which
corresponds to the contrast ratio at a mean intensity of 3 Mcps in
Fig.~\ref{ContrastIntensity}(b), where the measured results denoted
by circular dots show a linear decrease as the mean count rate
increases.

This linear decrease can be explained in terms of the correction
factor $\alpha$  of the detector which needs to be multiplied by the
measured intensity to get an original intensity,
\begin{equation}\label{correctionFactor}
    \alpha = \frac{I}{I_m}=1+T_d I,
\end{equation}
where $I$ and $I_m$ are the original and measured intensities,
respectively, in the unit of cps and $T_d$ is the dead time of the
detector.
If there exists additional dead times such as the dead time of the
counting board, $T_d$ should be replaced with an effective total
dead time including all dead times, and this subject will be
discussed in details in the next section. For $T_d a \ll 1$ and $T_d
b \ll 1$,
\begin{eqnarray}
    I_m(t) &\simeq& (1-T_d I(t))I(t) \simeq b-T_d b^2 + (a-2ab T_d)\sin\omega
    t \nonumber\\
    &\simeq& b(1-T_d b)\left[1+(a/b)(1-T_d b)\sin \omega t \right]\;,
\end{eqnarray}
which shows that the contrast ratio is modified from $a/b$ to
$(1-T_d b)(a/b)$.

Note that the unmodified contrast $a/b$, fixed in the experiment,
can be determined from the $y$ intercept of a linear fit to the
measured contrast ratios and the inclination of the linear fit
corresponds to the dead time $T_d$. In Fig\
\ref{ContrastIntensity}(b), we get $a/b\simeq 0.5$ and $T_d\simeq$
56 ns, which is about 10\% larger than our initial estimate based on
the single-photon pulse shape in Fig.\
\ref{DeadTimeDistribution}(b).
This discrepancy is due to the finite bin size of 12.5 ns, which
results in an additional broadening of about 6 ns, a half of the bin
size, in the effect of the detector dead time in the correlation
function.

\subsection{Dead Time Effect on $g^{(2)}(\tau)$}

\begin{figure}[b]\centering
\includegraphics[width=3.3in]{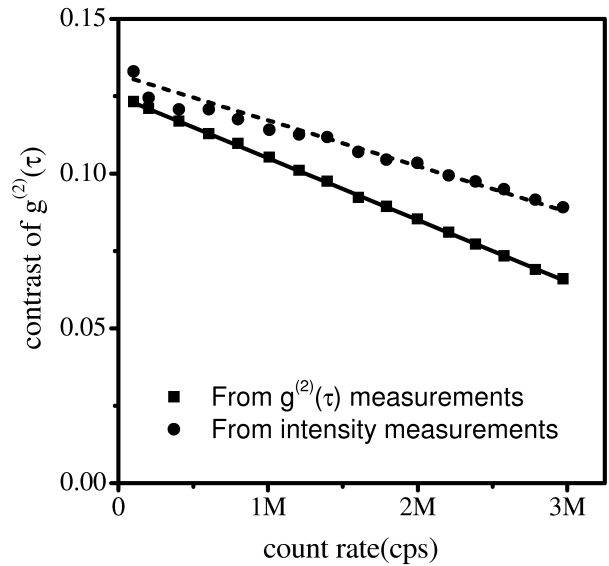}
\caption{Contrast of the second-order correlation function. Square
dots: calculated from the measured $g^{(2)}(\tau)$ (100 kHz),
circular dots: expected from the measured contrasts of the intensity
(0.1 Hz modulation). A solid and dashed lines are theoretical
results with the effective dead time of 80 ns and 56 ns,
respectively.} \label{g2contrast}
\end{figure}

Since the intensity profile is distorted by the detector dead time,
the second-order correlation function will be also affected. For a
measured intensity contrast of $a'/b'$, the expected contrast of the
second-order correlation function is $a'^2/2b'^2$. They are denoted
by circular dots in Fig.~\ref{g2contrast}. For an unmodified
contrast $a/b$, the dependence of the measured contrast $a'/b'$ on
the mean intensity $b$ is given by $(1-T_d b)a/b$. The contrast of
$g^{(2)}(\tau)$ is thus expected to be
\begin{equation}
a'^2/2b'^2 =\frac{1}{2}\left[(1-T_d b)(a/b)\right]^2\simeq (1-2T_d b)a^2/2b^2
\label{g2-contrast}
\end{equation}
The dashed line in Fig.~\ref{g2contrast} indicates the expected
contrast with $T_d=56$ ns.

Using MMTDC, we have measured the second-order correlation function
under the condition identical to the one under which we had measured
the intensity except for the intensity modulation frequency, which
is now set at 100kHz, since the correlation measurement need not be
done in a single-shot fashion as in the intensity measurement. The
square dots in Fig.~\ref{g2contrast} are the contrasts obtained by
fitting the measured $g^{(2)}(\tau)$ with a sinusoidal function. It
decreases linearly as the photon flux increases, but the decreasing
rate of the contrast ratio is 1.54 times larger than that expected
from the intensity measurement.

This discrepancy is due to the dead time of the counter/timing
board, which was absent in the preceding intensity measurement
performed at 0.1 Hz modulation frequency with a data transfer rate
of 100Hz. In the second-order correlation measurement, regardless of
the modulation frequency, since all of the arrival times are
recorded, the data transfer rate can be much faster than the inverse
of 250 ns, which is the maximum dead time of the counter. When the
time difference of two successive photons is shorter than 250 ns,
there exists a finite chance that the following photon is ignored. A
complexity arises since this chance is not alway unity. It can be
anywhere between 0 and 1. We thus need to find an effective dead
time which can properly includes both detector and counter dead time
effects.

\section{Analysis and Calibration Method}

\begin{figure}[t]\centering
\includegraphics[width=3.3in]{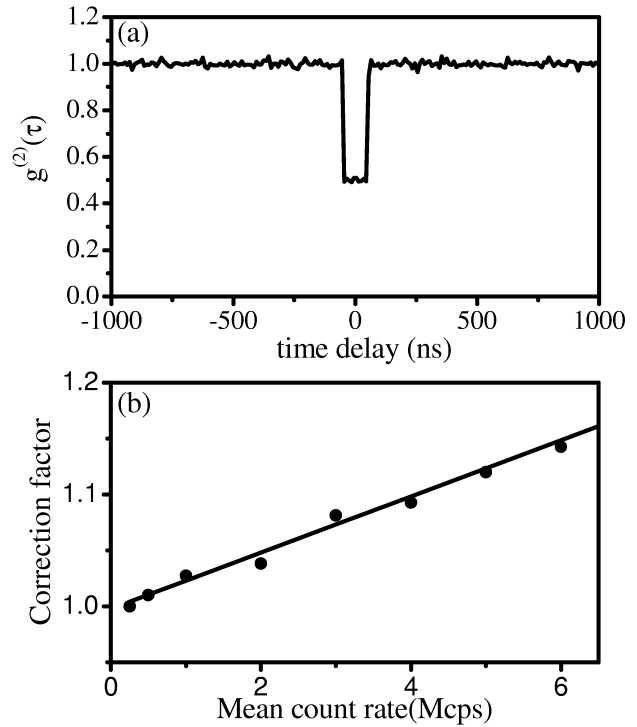}
\caption{Numerical simulation result for a partial dead time
effect. A dead time of 50 ns and a probability to lose photon of 0.5 are
assumed. (a) A second-order correlation function. (b) Correction
factors as a function of the mean count rate of incoming photons. Solid
line is a linear fit.}
\label{PartialDeadTime}
\end{figure}

To understand the dead time effect of counter, we numerically
simulated the effect of a partial dead time. A Poissonian light
source was simulated using a random number generation algorithm. In
the simulation, if the time difference between two successive
photons are shorter than 50 ns, the following photon is omitted with
a probability $P_L$ of 50\%. This probability would be unity for the
case of the detector dead time. Figure~\ref{PartialDeadTime}(a) is
the second-order correlation function of this simulated source and
shows a dip with a half width the same as the dead time and a depth
equal to $P_L$, 50\%. The correction factor $\alpha$ is calculated
as a function of the mean intensity and shown in
Fig.~\ref{PartialDeadTime}(b). The result can be well fitted by the
following relation.
\begin{equation}
    \alpha = \frac{I}{I_m}=1+P_L T_d I,
\end{equation}
Note that the linearity coefficient in $\alpha$ with respect to the
intensity is not $T_d$ as in Eq.~(\ref{correctionFactor}) but $P_L
T_d$. We can generalize this observation and expect that the
detector dead time $T_d$ in Eq.\ (\ref{g2-contrast}) will be
replaced by $P_L T_d$. This expectation has been confirmed by our
numerical simulations. Therefore, we call $T_{\rm eff}\equiv P_L
T_d$ an effective dead time and it is equal to the area of the dip
($\tau \ge 0$ only) around the origin in the second-order
correlation function obtained in the single-detector configuration
(Fig.~\ref{PartialDeadTime}(a)). In general, we can experimentally
determine the effective dead time $T_{\rm eff}^{\rm (tot)}$ of an
entire detection system from the dead time distribution, {\em i.e.},
the shape of the second-order correlation function around the
origin, calculated with photons detected on a single detector only.

The dead time distribution of our MMTDC system is shown in
Fig.~\ref{DeadTimeDistribution}(a). The total area of the dip near
the origin is 80 ns, which is our effective dead time $T_{\rm
eff}^{\rm (tot)}$ for the entire measurement system.  The area of
the central dip only equals $T_d$=55 ns, originating from the
detector dead time. The ratio $T_{\rm eff}^{\rm (tot)}/T_d$ is 1.45,
and thus we expect that the inclination of the observed contrast in $g^{(2)}(\tau)$
with respect to the counting rate $b$ would be 1.45 times greater
than that expected from the intensity measurement free from the
counter dead time effect. This is what we have observed in Fig\
\ref{g2contrast}, where the observed ratio is 1.54, showing only 6\%
discrepancy from our expectation.

From these observations, we can establish a calibration method for
the second-order correlation function of a non-stationary light
source. We first measure the second-order correlation function using
the single detector configuration and determine the effective dead
time $T_{\rm eff}^{\rm (tot)}$ of the entire system from the dead
time distribution around the origin. The second-order correlation
function is then measured again, but this time in the two detector
configuration. Since the correction factor in Eq.\ (6) is
independent of the modulation frequency, we can calibrate the
measured correlation function by deviding its amplitude by $(1- 2
T_{\rm eff}^{\rm (tot)}b)$, where $b$ is the mean count rate.



This work was supported by the Ministry of Science and Technology of Korea and NSF grant no. 9876974-PHY.



\begin{references}
\bibitem{HBT}
E.\ Hanbury Brown and R.\ Q.\ Twiss, Nature {\bf 177}, 27 (1957)

\bibitem{Morgan1966}
B.\ L.\ Morgan, L.\ Mandel, Phys.\ Rev.\ Lett.\ {\bf 16}, 1012
(1966)

\bibitem{Cummins1970}
H.\ Z.\ Cummins and H.\ L.\ Swinney, p.133 in {\it Progress in
Optics}, Vol. 8, ed. E.\ Wolf (North-Holland, Amsterdam)(1970)

\bibitem{Arecchi1966}
F.\ T.\ Arecchi, E.\ Gatti and A.\ Sona, Phys.\ Lett.\ {\bf 20}, 27
(1966)

\bibitem{Kimble-PRL77}
H.\ J.\ Kimble, M.\ Dagenais, and L.\ Mandel, Phys.\ Rev.\ Lett.\
{\bf39}, 691 (1977).


\bibitem{Short-PRL83}
R.\ Short and L.\ Mandel, Phys.\ Rev.\ Lett.\ {\bf 51},384 (1983).

\bibitem{Scarl1966}
D.\ B.\  Scarl, Phys.\ Rev.\ Lett.\ {\bf 17}, 663 (1966)

\bibitem{Phillips1967}
D.\ T.\ Phillips, H.\ Kleiman and S.\ P.\ Davis, Phys.\ Rev.\ {\bf
153} 113 (1967)

\bibitem{Davidson1968}
F.\ Davidson and L.\ Mandel, J.\ Appl.\ Phys.\ {\bf 39} 62 (1968)

\bibitem{Davidson1969}
F.\ Davidson, Phys.\ Rev.\ {\bf 185} 446 (1969)

\bibitem{Cummins1977}
H.\ Z.\ Cummins and E.\ R.\ Pike, {\it Photon Correlation
Spectroscopy and Velocimetry}, Plenum Press, New York (1977)

\bibitem{Swinney1983}
H.\ L.\ Swinney, Physica D {\bf 7} 3 (1983)

\bibitem{Pike1986}
E.\ R.\ Pike, in {\it Coherence, Cooperation and Fluctuations},
Cambridge University Press, Cambridge p. 293 (1986)

\bibitem{Mandel-OL79}
L.\ Mandel, Opt.\ Lett.\ {\bf 4}, 205 (1979).

\bibitem{An-PRL94}
K.\ An, J.\ J.\ Childs, R.\ R.\ Dasari, and M.\ S.\ Feld, Phys.\
Rev.\ Lett.\ {\bf 73}, 3375(1994).

\bibitem{Wonshik04}
W.\ Choi et al, Observation of noncalssical photon statistics in the
cavity-QED microlaser, to be submitted to the Phys.\ Rev.\ Lett.\
(2005)

\bibitem{Meschede-PRL85}
D.\ Meschede and H.\ Walther, and G.\ Muller, Phys.\ Rev.\ Lett.\
{\bf 54}, 551(1985).



\end{references}
\end{document}